\documentclass[showpacs,preprintnumbers,aps,10pt,amsmath,amssymb,prb]{revtex4}

\usepackage{graphicx}
\usepackage{subfigure}
\usepackage{dcolumn}
\usepackage{bm}

\newcommand{\kk}[0]{{\bf k}}
\newcommand{\q}[0]{{\bf q}}
\newcommand{\0}[0]{{\bf 0}}

\newcommand{\kko}[0]{{{\bf k}+{\bf k}_0}}
\newcommand{\ko}[0]{{{\bf k}_0}}

\newcommand{\ms}[0]{{$\alpha$-MnS}}
\newcommand{\me}[0]{{$\alpha$-MnSe}}

\begin{document}
\title{Order-by-disorder effects in antiferromagnets on face-centered cubic lattice}

\author{L.\ A.\ Batalov$^1$}
\email{zlokor88@gmail.com}
\author{A.\ V.\ Syromyatnikov$^{1,2}$}
\email{asyromyatnikov@yandex.ru}
\affiliation{$^1$National Research Center "Kurchatov Institute" B.P.\ Konstantinov Petersburg Nuclear Physics Institute, Gatchina 188300, Russia}
\affiliation{$^2$Saint Petersburg State University, St.Petersburg State University, 7/9 Universitetskaya nab., St. Petersburg, 199034 Russia}

\date{\today}

\begin{abstract}

We discuss the role of quantum fluctuations in Heisenberg antiferromagnets on face-centered cubic lattice with small dipolar interaction in which the next-nearest-neighbor exchange coupling dominates over the nearest-neighbor one. It is well known that a collinear magnetic structure which contains (111) ferromagnetic planes arranged antiferromagnetically along one of the space diagonals of the cube is stabilized in this model via order-by-disorder mechanism. On the mean-field level, the dipolar interaction forces spins to lie within (111) planes. By considering $1/S$ corrections to the ground state energy, we demonstrate that quantum fluctuations lead to an anisotropy within (111) planes favoring three equivalent directions for the staggered magnetization (e.g., $[11\overline{2}]$, $[1\overline{2}1]$, and $[\overline{2}11]$ directions for (111) plane). Such in-plane anisotropy was obtained experimentally in related materials MnO, \ms, \me, EuTe, and EuSe. We find that the order-by-disorder mechanism can contribute significantly to the value of the in-plane anisotropy in EuTe. Magnon spectrum is also derived in the first order in $1/S$.

\end{abstract}


\maketitle

\section{Introduction}

Frustrated spin systems have attracted a great deal of interest in recent years. \cite{balents} In many of them, classical ground state has a degeneracy which can be lifted by quantum or thermal fluctuations who thereby select and stabilize an ordered state. This is the so-called ``order by disorder'' phenomenon. \cite{vill,henley1,shender2} One of such spin systems is the Heisenberg antiferromagnet (AF) on face-centered cubic (fcc) lattice in which the next-nearest-neighbor AF exchange coupling (i.e., that along the cube edge) dominates over the nearest-neighbor one. \cite{shender} Although this model describes a number of prototypical AFs (e.g., MnO), some open problems remain in this field.

\begin{figure}
\noindent
\includegraphics[scale=0.2]{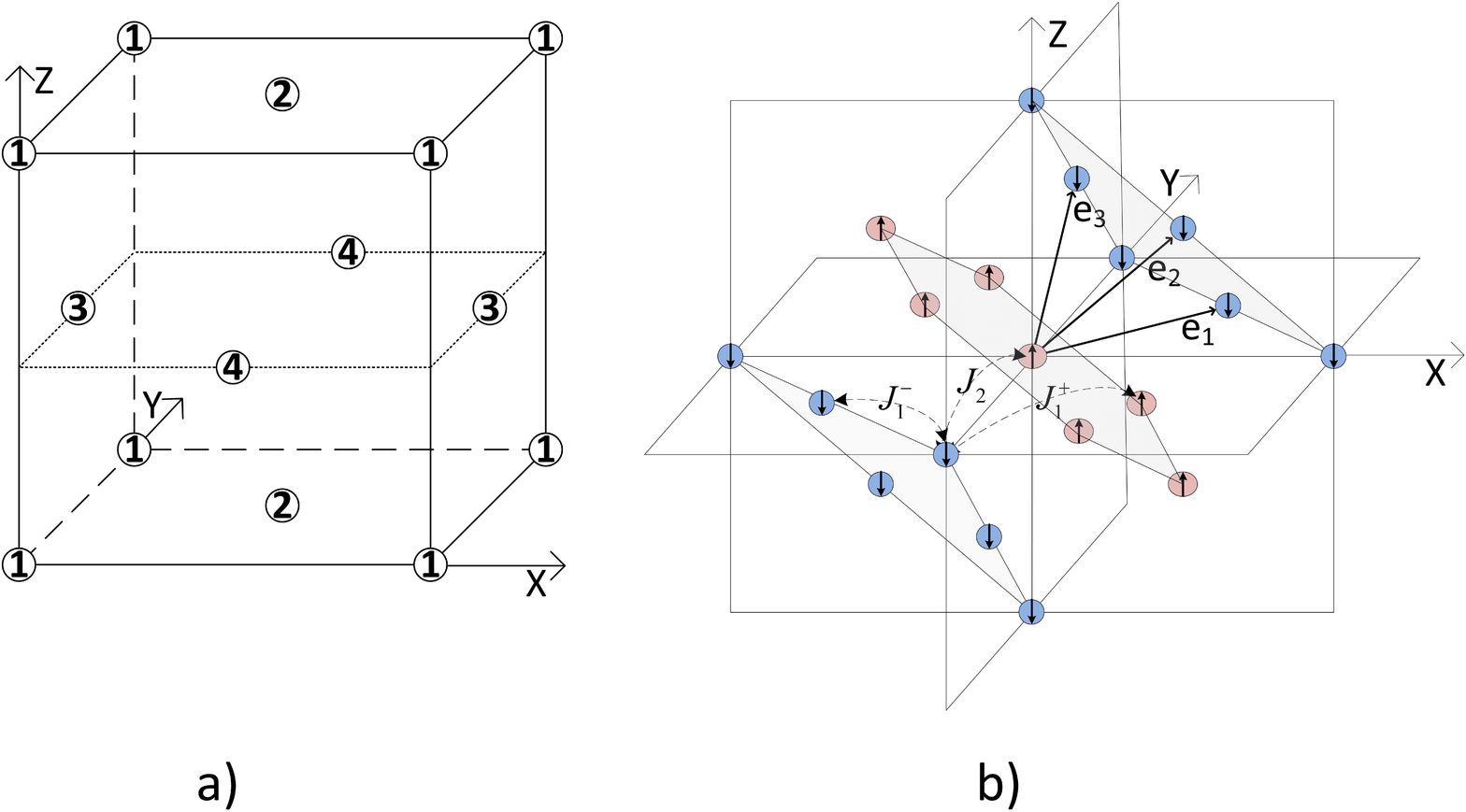}
\hfil
\caption{(Color online.) (a) Classical AF on fcc lattice at $T=0$. Number indicates which of four AF cubic sublattices the given spin belongs to. (b) Magnetic structure of AF on fcc lattice which is realized in the considered model via order-by-disorder mechanism (which is referred to in the literature as AF structure of the second kind, type $A$). Spins belonging to different (111) FM planes (shaded) arranged antiferromagnetically along $[111]$ direction are shown in different colors and are denoted by arrows $\uparrow$ and $\downarrow$. Exchange coupling constants $J_1^\pm$, $J_2$ and lattice vectors ${\bf e}_{1,2,3}$ are also presented.
\label{structure}}
\end{figure}

AF on fcc lattice can be viewed as four interpenetrating AF cubic sublattices (see Fig.~\ref{structure}(a)). \cite{shender,japan} Any spin from a sublattice locates at zero molecular field of spins from three other sublattices. As a result, staggered magnetizations of these sublattices can be oriented arbitrary relative to each other that leads to an infinite ground state degeneracy. However, quantum fluctuations make staggered magnetizations of all sublattices parallel to each other. \cite{shender} Besides, among two possible collinear arrangements, they select that presented in Fig.~\ref{structure}(b) which is referred to in the literature as AF structure of the second kind, type $A$ (fluctuations make unfavorable type $B$ structure). \cite{shender,japan} This AF structure contains (111) ferromagnetic (FM) planes arranged antiferromagnetically along one of $\langle1 1 1\rangle$ directions. As soon as $[111]$, $[\bar1 1 1]$, $[1 \bar1 1]$, and $[1 1 \bar1]$ directions are equivalent, there are four equivalent spin arrangements of this type which are described by vectors of the magnetic structure $\ko=(\pi,\pi,\pi)$, $(\pi,0,0)$, $(0,\pi,0)$, and $(0,0,\pi)$ (hereafter we set to unity the cube edge length). This symmetry breaking by fluctuations is naturally accompanied by appearance of gaps induced by fluctuations in some magnon branches (not all the magnon branches acquire gaps because the continuous symmetry remains related to a rotation of all spins by any angle about any axis). \cite{shender} It can be shown also that the selection of collinear spin structures can be described phenomenologically on the mean-field level by introducing to the Hamiltonian a biquadratic interaction between spins from different sublattices having the form $-Q({\bf S}_i{\bf S}_j)^2$, where $Q>0$. \cite{shender,henley2}

Unfortunately, it is often difficult to confirm unambiguously the presence of order-by-disorder effects in real materials because small anisotropic interactions cannot be generally excluded which are able to lift the classical ground state degeneracy explicitly (see also discussion in Ref.~\cite{bal}). (In particular, one always expects the biquadratic exchange in real substances bearing in mind that it arises naturally in the Hubbard model in high orders in $t/U$, where $t$ is the hoping constant and $U$ is the on-site repulsion energy. \cite{yosida}) However, there are some compounds in which low-symmetry interactions are ruled out \cite{bal} or they are strongly suppressed for some reason. Order-by-disorder effects contribute noticeably to properties of such materials. 

One expects this situation in the following AFs on fcc lattice which show the spin structure presented in Fig.~\ref{structure}(b) at small $T$: MnO, \cite{neu_mno,good} \ms, \cite{neu_mns} \me, \cite{neu_mns} EuTe, \cite{eute_neu} and EuSe. \cite{Griessen19712219,PhysRevB.81.155213} Four types of domains were observed in these materials at small $T$ in which four equivalent $(111)$ planes are FM planes. As soon as magnetic ions $\rm Mn^{2+}$ and $\rm Eu^{2+}$ are in isotropic states characterized by zero orbital moment in these compounds, the anisotropy arisen from spin-orbit interaction is expected to be very small. The main source of anisotropy is the dipolar interaction in these materials. It was found in Refs.~\cite{cohen,kaplan,1957} that anisotropic corrections to the classical ground state energy from dipolar interaction make (111) planes to be easy planes in accordance with experimental results. Dipolar forces make also more favorable type $A$ AF structure of the second kind rather than the type $B$ one (see Ref.~\cite{1957} and references therein), as quantum fluctuations do. However, dipolar interaction does not select the collinear spin arrangement which is observed experimentally. In MnO, the selection of the collinear magnetic structure was attributed to dipolar anisotropy arisen due to small lattice rhombohedral distortion. \cite{1957} But such lattice distortions were not observed in other compounds. It should be noted also that the biquadratic exchange was suggested phenomenologically well before Ref.~\cite{shender} to explain the temperature dependence of the order parameter and some other experimental findings in MnO and EuTe. \cite{eute_neu,bic1,bic2,bic3} Particular estimations using equations from Ref.~\cite{shender} give for the value of the effective biquadratic interaction $Q\sim 0.1$~K and $10^{-3}$~K for MnO and EuTe, respectively. These values are of the same order of magnitude as those proposed in Refs.~\cite{eute_neu,bic1,bic2,bic3} for description of experimental data. Then, the order-by-disorder mechanism has a large impact on magnetic properties of these materials.

Experimental data show also that there is a small anisotropy within (111) planes of unknown origin in all materials mentioned above. In particular, directions $[11\overline{2}]$, $[1\overline{2}1]$, and $[\overline{2}11]$ are three equivalent easy axes within (111) plane. It is the aim of the present paper to demonstrate that the order-by-disorder mechanism contributes to this anisotropy. For this purpose, we consider the first $1/S$ correction to the ground state energy of AF on fcc lattice with dipolar forces. We find below that due to the dipolar interaction this correction is anisotropic and it contributes to the in-plane anisotropy. Actually, we extend the analysis of order-by-disorder phenomena carried out in Ref.~\cite{shender} by inclusion the small dipolar interaction in the model. The obtained values of the in-plane anisotropy (and values of the gap in the magnon spectrum related to the in-plane anisotropy) are compared with those measured experimentally in considered substances. We point out that the order-by-disorder mechanism can contribute noticeably to the anisotropy in EuTe.

The rest of the present paper is organized as follows. We discuss the Hamiltonian transformation and technique in Sec.~\ref{hamtr}. The classical magnon spectrum is analysed in Sec.~\ref{specclass}. The spectrum renormalization in the model without dipolar forces is discussed in Sec.~\ref{specwdip}. The order-by-disorder effects in the model with dipolar forces are considered in Sec.~\ref{odis}, where we derive an expression for the in-plane anisotropy. Sec.~\ref{exper} contains comparison with available experimental data in MnO, \ms, \me, and EuTe. A summary and our conclusion can be found in Sec.~\ref{conc}.

\section{Hamiltonian transformation}
\label{hamtr}

We discuss Heisenberg AF on fcc lattice with dipolar interaction whose Hamiltonian has the form
\begin{equation}
\label{ham}
\mathcal{H} = \frac{1}{2}\sum_{l\neq m}
\left(
J_{lm}\delta_{\alpha \beta} - Q_{lm}^{\alpha \beta}
\right)S_l^\alpha S_{m}^\beta,
\end{equation}	
where the summation over repeated Greek letters is implied,
\begin{eqnarray}
\label{dip_forces}
Q_{lm}^{\alpha \beta} &=& \frac{\omega_0}{4\pi}\frac{3R_{lm}^\alpha R_{lm}^\beta-\delta_{\alpha \beta}R_{lm}^2}{R_{lm}^5},\\
\label{w0}
\omega_0 &=& 4\pi \frac{(g\mu_B)^2}{v_0},
\end{eqnarray}
$v_0$ is the unit cell volume, and non-zero exchange constants are
\begin{equation}
\label{exch}
J_{lm}=
\left\{
\begin{aligned}
& J_1^+, & |{\bf R}_l-{\bf R}_m| = \frac{1}{\sqrt{2}}, & \quad \langle {\bf S}_l\rangle \uparrow \downarrow \langle {\bf S}_m\rangle,
  \\
& J_1^-, & |{\bf R}_l-{\bf R}_m| = \frac{1}{\sqrt{2}}, & \quad \langle {\bf S}_l\rangle \uparrow \uparrow \langle {\bf S}_m\rangle,
  \\  
& J_2, & |{\bf R}_l-{\bf R}_m|=1, & { }
\end{aligned}
\right.
\end{equation}
where the length of the cube edge is set to be equal to unity, $\langle {\bf S}_l\rangle \uparrow \downarrow \langle {\bf S}_m\rangle$ and $\langle {\bf S}_l\rangle \uparrow \uparrow \langle {\bf S}_m\rangle$ denote that average magnetic moments at sites $l$ and $m$ are antiparallel and parallel to each other, respectively (see Fig.~\ref{structure}(b)). Two slightly different values of the nearest-neighbor exchange constants $J_1^+$ and $J_1^-$ in Eq.~\eqref{exch} arise in MnO due to lattice rhombohedral distortion \cite{Pepy1974433,PhysRevB.1.236} whereas $J_1^+=J_1^-=J_1$ in \ms, \me, EuTe, and EuSe. The exchange coupling along cube edges (i.e., between next-nearest-neighbor spins) is characterized by constant $J_2$.

By taking the Fourier transformation, one has from Eq. \eqref{ham}
\begin{equation}
\mathcal{H} = \frac{1}{2}\sum_\kk
\left(  J_\kk\delta_{\alpha \beta }-Q_\kk^{\alpha \beta}   \right)
S_\kk^\alpha S_{-\kk}^\beta , \label{ham_fourier}
\end{equation}
where $Q_\kk^{\alpha \beta}=\sum_l Q_{l m}^{\alpha \beta}\exp(i\kk {\bf R}_{l m})$ and $J_\kk=\sum_l J_{l m}\exp(i\kk {\bf R}_{l m})$. We use below the coordinate system with basis vectors ${\bf e}_1=(1/2,1/2,0)$, ${\bf e}_2=(1/2,0,1/2)$, and ${\bf e}_3=(0,1/2,1/2)$ (see Fig.~\ref{structure}(b)) so that ${\bf R}_n=n_1{\bf e}_1+n_2{\bf e}_2+n_3{\bf e}_3$, where $n_{1,2,3}$ are integer. The corresponding basis vectors of the reciprocal lattice are given by ${\bf b}_1=(1,1,-1)$, ${\bf b}_2=(1,-1,1)$, and ${\bf b}_3=(-1,1,1)$ so that the magnon momentum has the form ${\bf k}=k_1{\bf b}_1+k_2{\bf b}_2+k_3{\bf b}_3$, where $k_{1,2,3}\in(-\pi,\pi]$. One obtains in this case
\begin{equation}
	J_\kk = J_{1\kk}^+ + J_{1\kk}^- + J_{2\kk},
\end{equation}
where
\begin{eqnarray}
\label{jjj}
J_{1\kk}^+ &=& 2J_1^+\left( \cos k_1 + \cos k_2 + \cos k_3 \right),\nonumber\\
J_{1\kk}^- &=& 2J_1^-\left( \cos(k_1-k_2) + \cos(k_1-k_3) + \cos(k_2-k_3) \right),\\
J_{2\kk} &=& 2J_2\left( \cos(k_1+k_2-k_3) + \cos(k_1-k_2+k_3) + \cos(k_1-k_2-k_3) \right).\nonumber
\end{eqnarray}

It is convenient to represent spin components as follows: 
${\bf S}_l=S_l^x\hat{x}+(S_l^y\hat{y}+S_l^z\hat{z})\exp(i\kk_0{\bf R}_l)$, where $\hat{x}$, $\hat{y}$, and $\hat{z}$ are mutually orthogonal unit vectors which can be directed arbitrarily relative to cube edges, $\ko$ is a vector describing an AF magnetic structure like that shown in Fig.~\ref{structure}(b), and $\exp(i\kk_0{\bf R}_l)$ is equal to $+1$ and $-1$ when $\langle {\bf S}_l\rangle$ is parallel and antiparallel to $\langle {\bf S}_0\rangle$, correspondingly. The particular forms of $\ko$ describing four different AF structures with ferromagnetic (111) planes are $(\pi,\pi,\pi)$, $(\pi,0,0)$, $(0,\pi,0)$, and $(0,0,\pi)$. We use below the Holstein-Primakoff spin representation having the form
\begin{eqnarray}
S_l^x &\approx& \sqrt{\frac{S}{2}}\left(  a_l+a_l^\dagger-\frac{a^\dagger_l a_l^2 + (a^\dagger_l)^2 a_l}{4S}
\right), \nonumber\\
S_l^y &\approx& -i\sqrt{\frac{S}{2}}\left(   a_l-a_l^\dagger-\frac{a^\dagger_l a_l^2 - (a^\dagger_l)^2 a_l}{4S}
\right), \label{dm}\\
S_l^z &=& S-a^\dagger_l a_l. \nonumber
\end{eqnarray}
Although it is implied in Eq.~\eqref{dm} that the spin value $S$ is much greater than unity, it is well known that first $1/S$ corrections make major contributions to observable quantities even at $S\sim1$ in nonfrustrated systems far from critical points. That is why we restrict ourselves below by analysis of only first $1/S$ corrections. Besides, the particular compounds we discuss below have quite large $S$ (5/2 or 7/2).

Taking the Fourier transformation in Eqs.~\eqref{dm} and using the relation ${\bf S}_\kk = S_\kk^x\hat{x}+S_\kko^y\hat{y}+S_\kko^z\hat{z}$ one obtains from Eq.~\eqref{ham_fourier} for the Hamiltonian $\mathcal{H}=E_{cl}+\sum_{i=1}^6\mathcal{H}_i$, where 
\begin{equation}
\label{e0}
\frac 1N E_{cl} = -6 S^2(J_2+J_1^+-J_1^-) -\frac12 S^2 Q_\ko^{zz}
\end{equation}
is the classical ground-state energy, $N$ is the number of spins in the lattice, and $\mathcal{H}_i$ denote terms containing products of $i$ operators $a^\dagger$ and $a$. In particular, one has
\begin{eqnarray}
\label{h1}
\frac{1}{\sqrt N}\mathcal{H}_1 &=& -S\sqrt{\frac{S}{2}}Q^{xz}_\ko(a_\ko+a^\dagger_\ko)+i S\sqrt{\frac{S}{2}}Q^{yz}_\ko(a_\0-a^\dagger_\0),\\
\label{h2}
\mathcal{H}_2 &=& \sum_{\bf k}
\left( 
E_{\bf k}a^\dagger_{\bf k}a^{}_{\bf k} + \frac{B_{\bf k}}{2}\left(a^{}_{\bf k}a^{}_{-{\bf k}}+a^\dagger_{\bf k}a^\dagger_{-{\bf k}}\right)
+\mathcal{E}_{\bf k}a^\dagger_{{\bf k}+{\bf k}_0}a^{}_{\bf k}
+ \frac{\mathcal{B}_{\bf k}}{2} a^{}_{{\bf k}}a^{}_{-{\bf k}+{\bf k}_0}
+ \frac{\mathcal{B}_{\bf k}^*}{2} a^\dagger_{{\bf k}}a^\dagger_{-{\bf k}+{\bf k}_0}
\right),\\
\label{h2_denote_e}
E_\kk & = &  \frac{S}2\left( J_\kk+J_\kko-2J_\ko \right) - \frac{S}{2}\left(Q^{xx}_\kk+Q^{yy}_\kko-2Q^{zz}_\ko \right),\nonumber\\
B_\kk & = &  \frac{S}2\left( J_\kk-J_\kko \right)  - \frac{S}{2}\left(Q^{xx}_\kk-Q^{yy}_\kko\right),\\
\mathcal{E}_\kk & = & i\frac{S}{2}\left( Q_\kko^{xy}-Q^{xy}_\kk\right),\nonumber\\
\mathcal{B}_\kk & = & i\frac S2 \left( Q_\kko^{xy}+Q^{xy}_\kk\right).\nonumber
\end{eqnarray}

Notice that the dipolar tensor components in Eqs.~\eqref{e0}--\eqref{h2_denote_e} are calculated in the coordinate system in which $z$ axis is directed along the staggered magnetization. Then, the ground state energy depends on the direction of the staggered magnetization relative to cube edges. To find the direction of the staggered magnetization which minimizes the free energy, we express dipolar tensor components $Q^{\alpha\beta}_\kk$  in Eqs.~\eqref{e0}--\eqref{h2_denote_e} as linear combinations of components $\widetilde Q^{\alpha\beta}_\kk$ calculated in the coordinate system which axes are parallel to cube edges. Introducing three direction cosines $\gamma_x$, $\gamma_y$, and $\gamma_z$ of the staggered magnetization with respect to cube edges, one obtains from Eq.~\eqref{e0} for the anisotropic part of the classical ground state energy
\begin{equation}
\label{e_gs_an}
E_{cl}^{anis} = -\frac12 S^2 N
( \widetilde Q_\ko^{xx}\gamma_x^2 + \widetilde Q_\ko^{yy}\gamma_y^2 + \widetilde Q_\ko^{zz}\gamma_z^2 + 2\widetilde Q^{xy}_\ko\gamma_x\gamma_y + 2\widetilde Q^{xz}_\ko\gamma_x\gamma_z + 2\widetilde Q^{yz}_\ko\gamma_y\gamma_z).
\end{equation}
Hereafter we assume for definiteness that the AF ordering is characterized by vector $\ko=(\pi,\pi,\pi)$ (it is the magnetic structure which is presented in Fig.~\ref{structure}(b)). Particular calculations show that 
\begin{equation}
\label{qk0}
\begin{array}{l}
\widetilde Q_\ko^{xx} = \widetilde Q_\ko^{yy} = \widetilde Q_\ko^{zz}=0,\\
\widetilde Q_\ko^{xy} = \widetilde Q_\ko^{yz} = \widetilde Q_\ko^{xz}=-\eta\omega_0
\end{array}
\end{equation}
at $\ko=(\pi,\pi,\pi)$ and one has from Eq.~\eqref{e_gs_an}
\begin{eqnarray}
\label{ean_k}
E_{cl}^{anis} &=& N K_1(\gamma_x\gamma_y+\gamma_x\gamma_z+\gamma_y\gamma_z),\\
\label{k1}
K_1 &=& \eta S^2 \omega_0,\\
\label{eta}
\eta &\approx& 0.288.
\end{eqnarray}
Taking into account the obvious identity $(\gamma_x+\gamma_y+\gamma_z)^2=1+2(\gamma_x\gamma_y+\gamma_x\gamma_z+\gamma_y\gamma_z)$, we obtain from Eq.~\eqref{ean_k} that the following condition should hold to minimize the classical energy \eqref{ean_k}:
\begin{equation}
\label{111}
\gamma_x+\gamma_y+\gamma_z=0.
\end{equation}
This condition is fulfilled if the staggered magnetization lies within (111) plane. This result was found previously theoretically in the discussed model. \cite{cohen} Such spin arrangement was observed also experimentally in all considered materials. \cite{neu_mno,neu_mns,Zinn197623} Notice that $\mathcal{H}_1$ given by Eq.~\eqref{h1} vanishes if Eq.~\eqref{111} holds. As soon as the rotational invariance is preserved within (111) planes, one expects at least one gapless branch in the magnon spectrum in the spin-wave approximation (i.e., in the classical magnon spectrum). 

\section{Classical magnon spectrum}
\label{specclass}

Analysis of the bilinear part of the Hamiltonian \eqref{h2} can be carried out in a standard way as it is done, e.g., in Ref.~\cite{our}. This analysis shows that dipolar forces split the magnon spectrum into two branches which energies have the form in the spin waves approximation
\begin{equation}
\label{specsw}
\left(\epsilon_{\kk}^\pm\right)^2 =\left(\epsilon_{\kko}^\pm\right)^2 =A_\kk+A_\kko \pm \sqrt{(A_\kk+A_\kko)^2 - F_\kk F_\kko},
\end{equation}
where
\begin{eqnarray}
\label{af}
\frac{2}{S^2} A_\kk &=& 
(J_\kk-J_\ko+Q^{zz}_\ko+Q^{zz}_\kk)(J_\kko-J_\ko+Q^{zz}_\ko)
+Q^{xx}_\kk Q^{yy}_\kko - Q^{xy}_\kk Q^{xy}_\kko,\nonumber\\
\frac{1}{S^2} F_\kk &=& 
(J_\kk-J_\ko+Q^{zz}_\ko+Q^{zz}_\kk)(J_\kk-J_\ko+Q^{zz}_\ko)
+ Q^{xx}_\kk Q^{yy}_\kk-(Q^{xy}_\kk)^2,
\end{eqnarray}
$Q^{zz}_\ko=\eta\omega_0$ when condition \eqref{111} holds, and $\eta$ is given by Eq.~\eqref{eta}. It is seen from Eqs.~\eqref{af} that the square root vanishes in Eq.~\eqref{specsw} at $\omega_0=0$ and there is only one magnon branch without the dipolar interaction. As soon as $\epsilon_{\kk}^\pm$ are invariant under replacement of $\kk$ by $\kko$, one can consider only the neighborhood of the point $\bf k=0$ discussing long-wavelength magnons. It can be shown that $F_\kk=0$ at $\kk=\ko$ if Eq.~\eqref{111} holds. Then, one concludes from Eq.~\eqref{specsw} that $\epsilon_{\kk}^-$ and $\epsilon_{\kk}^+$ are gapless and gapped branches, respectively, in the spin-wave approximation. Eq.~\eqref{specsw} gives at $k\ll 1$ and $\omega_0\ll |J_1^\pm|,J_2$
\begin{eqnarray}
\label{e-}
\epsilon_{\bf k}^- &=& D(\theta_\kk,\phi_\kk)k,\\
\label{e+}
\epsilon_{\bf k}^+ &=& \sqrt{ D(\theta_\kk,\phi_\kk)^2k^2 + \Delta_+^2},
\end{eqnarray}
where 
\begin{eqnarray}
\label{Dd}
D(\theta_\kk,\phi_\kk)^2 &=& 3S^2(J_1^++J_2)
\left(
4J_2-2J_1^-+2J_1^+
+(\sin2\theta_\kk (\sin\phi_\kk + \cos\phi_\kk) + \sin^2\theta_\kk\sin2\phi_\kk)(J_1^++J_1^-)
\right),\\
\label{d+}
\Delta_+^2 &=& 36S^2(J_1^++J_2)\eta\omega_0,   
\end{eqnarray}
$\theta_\kk$ and $\phi_\kk$ are polar and azimuthal angles of $\kk$ in the coordinate system which axes are parallel to cube edges. 

It can be shown also that $F_\kk=0$ at $\kk=(\pi,0,0)$, $(0,\pi,0)$, and $(0,0,\pi)$ if Eq.~\eqref{111} holds and $J_1^+=J_1^-$. Then, $\epsilon_{\kk}^-$ given by Eq.~\eqref{specsw} vanishes at these three points too if $J_1^+=J_1^-$. 

Stability of spectra \eqref{e-} and \eqref{e+} requires $\Delta_+^2>0$ and $D(\theta_\kk,\phi_\kk)^2>0$. As it is seen from Eqs.~\eqref{Dd} and \eqref{d+}, the former condition satisfies at
\begin{equation}
\label{stab1}
J_1^++J_2>0	
\end{equation}
whereas the later one holds either at $J_1^-+J_1^+<0$ or at $J_1^-+J_1^+>0$ and
\begin{equation}
\label{stab2}
4J_2-3J_1^-+J_1^+>0.
\end{equation}

It is quite natural that the gap ($\Delta_+$) in one of the magnon branches ($\epsilon_{\bf k}^+$) is accompanied by the easy-plane dipolar anisotropy \eqref{ean_k} in the classical ground state energy. The correspondence between the anisotropy and the gap is even quantitative. Indeed, one leads to Eq.~\eqref{d+} for the gap in one of the magnon branches considering the Heisenberg AF on fcc lattice without dipolar forces and with one-ion anisotropy 
$
\left( 3\eta\omega_0/2 \right) \sum_i (S_i^\|)^2
$ 
(which models anisotropy \eqref{ean_k}), where $S_i^\|$ is the projection of ${\bf S}_i$ on [111] direction.

\section{Magnon spectrum renormalization without dipolar forces}
\label{specwdip}

It is instructive to consider the first $1/S$ corrections to the magnon spectrum in the considered model neglecting the smallest interaction, i.e., dipolar forces. It is the model which is discussed in Ref.~\cite{shender}. However, we do not repeat here calculations of Ref.~\cite{shender} by dividing the lattice into four interpenetrating cubic AF sublattices. We rather use the main result of Ref.~\cite{shender} that quantum fluctuations stabilize the collinear sublattices arrangement and calculate the spectrum in the first order in $1/S$ in the collinear state. The classical magnon spectrum obtained from Eq.~\eqref{specsw} has the form
\begin{equation}
\label{spec0}
\epsilon_{\bf k} = \sqrt{E_\kk^2-B_\kk^2},
\end{equation}
where $E_\kk$ and $B_\kk$ are given by Eqs.~\eqref{h2_denote_e} at $\omega_0=0$. Notice that the classical spectrum vanishes at $\kk=\bf 0$, $(\pi,\pi,\pi)$, $(\pi,0,0)$, $(0,\pi,0)$, and $(0,0,\pi)$ if $J_1^+=J_1^-$. 

Corrections to the spectrum in the first order in $1/S$ can be found using the standard Hartree
decoupling of the fourth-order interaction terms in the Hamiltonian which have the form
\begin{equation}
\label{h4}
\mathcal{H}_4 = \frac{1}{8N}\sum_{{\bf k}_1+{\bf k}_2+{\bf k}_3+{\bf k}_4={\bf 0}}
\left( 
a_{{\bf k}_1}^\dagger a^{}_{-{\bf k}_2} a^{}_{-{\bf k}_3} a^{}_{-{\bf k}_4} 
\left(J_{{\bf k}_2+\ko} - J_{{\bf k}_2} \right)
+ a_{{\bf k}_1}^\dagger a_{{\bf k}_2}^\dagger a^{}_{-{\bf k}_3} a^{}_{-{\bf k}_4} 
\left(2 J_{{\bf k}_1+{\bf k}_3+\ko} - J_{{\bf k}_1+\ko} - J_{{\bf k}_1} \right)
\right)
+h.c.,	
\end{equation}
where $h.c.$ denote terms which are Hermitian conjugated to presented ones (notice that third-order interaction terms ${\cal H}_3$ vanish at $\omega_0=0$). As a result of this procedure, one leads to the following renormalization of $E_\kk$ and $B_\kk$:
\begin{eqnarray}
\label{eb}
E_\kk & = &  \frac{S-\delta S}2\left( J_\kk+J_\kko-2J_\ko \right)
+ f_1 J_{1\bf 0}^+ + f_2J_{2\bf 0} 
- g (J_{1\bf 0}^- - J_{1\bf k}^-),\nonumber\\
B_\kk & = &  \frac{S-\delta S}2\left( J_\kk-J_\kko \right)
+ f_1 J_{1\kk}^+ + f_2J_{2\kk},
\end{eqnarray}
where $J_{1\kk}^\pm$ and $J_{2\kk}$ are given by Eqs.~\eqref{jjj}, and
\begin{eqnarray}
f_1 &=& \frac1N \sum_{\bf q} \frac{B_{\bf q}}{2\epsilon_{\bf q}} \frac{J_{1\bf q}^+}{J_{1\bf 0}^+},
\qquad
f_2 = \frac1N \sum_{\bf q} \frac{B_{\bf q}}{2\epsilon_{\bf q}} \frac{J_{2\bf q}}{J_{2\bf 0}},\nonumber\\
g &=& \frac1N \sum_{\bf q} \frac{E_{\bf q}-\epsilon_{\bf q}}{2\epsilon_{\bf q}} \frac{J_{1\bf q}^-}{J_{1\bf 0}^-},
\qquad
\delta S = \frac1N \sum_{\bf q} \frac{E_{\bf q}-\epsilon_{\bf q}}{2\epsilon_{\bf q}} 
= \langle a_l^\dagger a^{}_l \rangle
\end{eqnarray}
are dimensionless constants which are normally much smaller than unity. Then, the spectrum renormalization is small in the whole Brillouin zone except for some special points. It can be easily shown using Eqs.~\eqref{spec0} and \eqref{eb} that the renormalized spectrum vanishes at $\kk=\bf 0$ and $\kk=(\pi,\pi,\pi)$ and it has the form at $\kk=(\pi,0,0)$, $(0,\pi,0)$, and $(0,0,\pi)$
\begin{equation}
\label{gapss}
\epsilon_{\bf k} = \sqrt{32S(3J_2+J_1^+-2J_1^-)(S(J_1^+-J_1^-)+J_1^+f_1-J_1^-g)}.
\end{equation}
It is seen from Eq.~\eqref{gapss} that fluctuations lead to gaps at $\kk=(\pi,0,0)$, $(0,\pi,0)$, and $(0,0,\pi)$ at $J_1^+=J_1^-$. These are gaps which are obtained in Ref.~\cite{shender} and which are related to the order-by-disorder effect discussed there (i.e., the collinearity of four AF sublattices induced by fluctuations). 

We demonstrate now that the order-by-disorder mechanism opens gaps also at $\kk=\bf 0$ and $\kk=(\pi,\pi,\pi)$ if one takes into account dipolar forces.

\section{Easy-plane and in-plane anisotropies}
\label{odis}

Let us derive quantum corrections to the ground state energy which lift the rotational invariance within (111) plane and which have the form in the first order in $1/S$
\begin{equation}
\label{an_H}
\langle \mathcal{H}_2  \rangle 
= 
\sum_\q 
\frac{\epsilon^+_\q + \epsilon^-_\q - 2E_\q }{4}.
\end{equation}
Considering $\omega_0$ as the smallest parameter in the system, we expand Eq.~\eqref{an_H} up to the third order in $\omega_0$. Terms of the first two orders in $\omega_0$ do not depend on $\gamma_{x,y,z}$ if condition \eqref{111} is fulfilled. One obtains after tedious calculations for the anisotropic part of the third-order term
\begin{eqnarray}
\label{anis}
\langle \mathcal{H}  \rangle_{anis} 
&=&
- N K_2 \cos^23\varphi,\\
\label{k2}
K_2 &=& \frac{S\omega_0^3}{J_2^2}C,\\
\label{c}
C
&=& 
\frac{J_2^2S^2}{576\omega_0^3}
\frac1N
\sum_\q \frac{\sqrt{E_\q-B_\q}}{(E_\q+B_\q)^{5/2}}
\left(
\frac43
(\widetilde Q^{xy}_\q + \widetilde Q^{xz}_\q - 2\widetilde Q^{yz}_\q)
(\widetilde Q^{xy}_\q + \widetilde Q^{yz}_\q - 2\widetilde Q^{xz}_\q)
(\widetilde Q^{xz}_\q + \widetilde Q^{yz}_\q - 2\widetilde Q^{xy}_\q)
\right. \nonumber\\
&&\left.
{}+6
(\widetilde Q^{xy}_\q - \widetilde Q^{xz}_\q)
(\widetilde Q^{xx}_\q - \widetilde Q^{yy}_\q)
(\widetilde Q^{xx}_\q - \widetilde Q^{yy}_\q + 2\widetilde Q^{xy}_\q - 2\widetilde Q^{xz}_\q)
-\frac92 \widetilde Q^{xx}_\q \widetilde Q^{yy}_\q \widetilde Q^{zz}_\q
\right),
\end{eqnarray}
where $E_\q$ and $B_\q$ are given by Eqs.~\eqref{eb} and $\varphi$ is the angle between the staggered magnetization and one of the directions within (111) plane: $[11\overline{2}]$, $[1\overline{2}1]$, or $[\overline{2}11]$. Particular numerical calculations show that the constant $C$ is positive when conditions of the spectrum stability \eqref{stab1} and \eqref{stab2} hold. Then, the minimum of the anisotropic correction \eqref{anis} is achieved when the staggered magnetization is directed along $[11\overline{2}]$, $[1\overline{2}1]$, or $[\overline{2}11]$ (easy directions).

Notice that we use Eqs.~\eqref{eb} for $E_\q$ and $B_\q$ in Eqs.~\eqref{c} rather than their bare values \eqref{h2_denote_e} at $\omega_0=0$. The reason for that is the following. Let us consider first the case of $J_1^+\ne J_1^-$ and $\omega_0\ll|J_1^--J_1^+|$. Neglecting quantum renormalization of $E_\q$ and $B_\q$, one finds $E_\q+B_\q=J_\q-J_\ko=0$ only at $\q=\ko$ so that the denominator of the summand in Eq.~\eqref{c} is proportional to $|\q-\ko|^5$ at $\q\sim\ko$. Taking into account Eqs.~\eqref{qk0}, one expects that numerator of the summand in Eq.~\eqref{c} is proportional to at least $|\q-\ko|^3$ at $\q\sim\ko$. Then, the singularity of the summand at $\q=\ko$ is summable and it is not required to take into account the small renormalization of $E_\q$ and $B_\q$ if $J_1^+\ne J_1^-$ and $\omega_0\ll|J_1^--J_1^+|$. However, the situation changes at $J_1^+ = J_1^- = J_1$ because three extra points arise at which $J_\q-J_\ko=0$: $\q=(\pi,0,0)$, $(0,\pi,0)$, and $(0,0,\pi)$ ($J_\q-J_\ko=8(J_1^+ - J_1^-)$ at such $\q$). It can be shown that $\widetilde{Q}_\q^{\alpha\alpha}=0$ at these points so that the last two terms in the brackets in Eq.~\eqref{c} are negligible in comparison with the first one that is equal to $64\eta^3\omega_0^3$ at such $\q$, where $\eta$ is given by Eq.~\eqref{eta}. As a result, singularities of the summand in Eq.~\eqref{c} are not summable at $\q=(\pi,0,0)$, $(0,\pi,0)$, and $(0,0,\pi)$ when $J_1^+ = J_1^-$. Then, we have to carry out calculations self-consistently by taking into account the renormalization of $E_\q$ and $B_\q$ that leads to gaps in the considered momenta screening the singularities in Eq.~\eqref{c}. Then, the range of validity of the expansion in powers of $\omega_0$ reads at $\omega_0\gg|J_1^--J_1^+|$ as $S\omega_0\ll|J_1^\pm|,J_2$. 

As soon as the rotational invariance is lifted in (111) planes by quantum fluctuations, one can naturally expect that quantum corrections of the first order in $1/S$ lead also to the gap in the gapless branch of the magnon spectrum. This expectation is realized in other Heisenberg magnets with dipolar forces. \cite{syromyat3d1,syromyat2d,syromyat3d,our} These previous calculations show that the gap in the magnon spectrum (but not the magnon damping) can be found in the spin-wave approximation by taking into account the anisotropy phenomenologically, i.e., by adding to the Hamiltonian a one-ion anisotropy modeling that of the fluctuation origin. In the present case, this effective anisotropy is obtained from Eq.~\eqref{anis} by expanding $\cos^23\varphi$ and replacing $\cos\varphi$ and $\sin\varphi$ by $S_i^z$ and $S_i^y$ (or $S_i^x$), respectively. As a result one has for the effective anisotropy
\begin{equation}
\frac{K_2}{2S^6}
\sum_i 
\left(
\left(S_i^y\right)^6 
+15 \left(S_i^z\right)^4\left(S_i^y\right)^2 
-15 \left(S_i^z\right)^2\left(S_i^y\right)^4
-\left(S_i^z\right)^6 
\right).
\end{equation}
This term leads to the following renormalization of coefficients $E_\kk$ and $B_\kk$ in the bilinear part of the Hamiltonian \eqref{h2}: 
\begin{equation}
\label{ren}
\begin{array}{l}
E_\kk\rightarrow E_\kk + 21K_2/2S,\\	
B_\kk\rightarrow B_\kk-15K_2/2S.
\end{array}
\end{equation}
As a result of this renormalization, one obtains for the gap square in the branch $\epsilon_\kk^-$
\begin{equation}
\label{d-}
\Delta_-^2 = 36 K_2 (J_1^++J_2) = 36\frac{S\omega_0^3(J_1^++J_2)}{J_2^2}C.
\end{equation}

It is important also to take into account quantum correction to the value $K_1$ of the easy-plane dipolar anisotropy \eqref{k1} which acquires the form in the first order in $1/S$ and in the leading order in $\omega_0$
\begin{equation}
\label{k1cor}
K_1 = \eta S^2 \omega_0 
- 
\frac S2\frac1N
\sum_\q 
\left(
1 - \sqrt{ \frac{J_{\q+\ko}-J_\ko}{J_\q-J_\ko} } 
\right)
\left( 2\widetilde{Q}_\ko^{xy} + \widetilde{Q}_\kk^{xy} \right).
\end{equation}
The gap $\Delta_+$ given by Eq.~\eqref{d+} is renormalized accordingly (see discussion at the end of Sec.~\ref{specclass}):
\begin{equation}
\label{d+cor}
\Delta_+^2 = 36(J_1^++J_2)K_1, 
\end{equation}
where $K_1$ is given by Eq.~\eqref{k1cor}. Quantum fluctuations reduce $K_1$ and $\Delta_+$ and give noticeable corrections to these quantities which are important for description of experimental results (see below). While classical expressions \eqref{k1} and \eqref{d+} for $K_1$ and $\Delta_+$ are well known, we are not aware of a discussion of their quantum renormalization. Then, Eqs.~\eqref{k1cor} and \eqref{d+cor} is an extension of the previous analysis.

\section{Comparison with experiment}
\label{exper}

Values of gaps $\Delta_+$ and $\Delta_-$ obtained theoretically using Eqs.~\eqref{d-} and \eqref{d+cor} and measured experimentally are summarized in the Table. It is seen from the Table that in accordance with previous results $\Delta_+$ is determined by the easy-plane anisotropy of the dipolar origin. One can see also that quantum corrections to the easy-plane anisotropy and to $\Delta_+$ given by Eqs.~\eqref{k1cor} and \eqref{d+cor} move theoretical results closer to experimental ones (except for \me). 

We have calculated $\Delta_-$ using Eq.~\eqref{d-} with constants $C$ obtained in two ways (the last two columns in the Table): by expanding Eq.~\eqref{an_H} to the third order in $\omega_0$ (i.e., using Eq.~\eqref{c}) and from the analysis of Eq.~\eqref{an_H} not using the $\omega_0$-expansion. It is seen from the Table that the $\omega_0$-expansion does not work in EuTe and EuSe because exchange constants are comparable with $\omega_0$ in these substances.

Unfortunately, in contrast to $\Delta_+$, values of $\Delta_-$ obtained using different experimental data deviate significantly from each other. Ranges are presented in the Table within which all values of $\Delta_-$ lie. Calculations of $\Delta_-$ using Eqs.~\eqref{c} and \eqref{d-} demonstrate that the dipolar contribution to $\Delta_-$ is much smaller than values obtained experimentally in MnO, \ms, and \me. Seemingly, another mechanism is mainly responsible for the anisotropy in (111) planes in these materials.

The situation is different in EuTe ($S=7/2$, $J_1^+=J_1^-=-0.384$~K, $J_2=0.626$~K, $\omega_0=0.44$~K, $T_N=9.6$~K) \cite{EuTe_NMR,PhysRevB.1.3021,PhysRevB.5.2634,PhysRevB.81.155213}. The phenomenological expression for the gap in lower magnon branch which was used in previous studies has the form $\Delta_-=\sqrt{2H_aH_E}$, where $H_E=6S(J_1^++J_2)$ is the exchange field which was found to be approximately 37~kOe (see Refs.~\cite{EuTe_NMR,PhysRevB.1.3021,PhysRevB.5.2634} and references therein) and $H_a$ is the effective field of the in-plane anisotropy. The following values were obtained experimentally for $H_a$: $8\pm4$~Oe (Ref.~\cite{PhysRevB.1.3021}), $12.4\pm1.4$~Oe (Ref.~\cite{EuTe_NMR}), $15\pm1$~Oe (Ref.~\cite{demo}), and $19.2\pm3$~Oe (Ref.~\cite{EuTe_NMR}) at $T=1.17$~K, $1.8$~K, 2~K, and 4.2~K, respectively. These data indicate that $\Delta_-$ lies in the range $0.06\div0.17$~K. Our calculations based on the analysis of Eq.~\eqref{an_H} and on Eq.~\eqref{d-} lead to $\Delta_- \approx 0.03$~K. Then, one can expect that the order-by-disorder mechanism contributes noticeably to the in-plane anisotropy in EuTe. However, further experimental activity is needed to find a more precise value of $K_2$ in this compound. It would be interesting also to compare our results with experimental ones in another relevant material containing Eu, EuSe, but we are not aware of corresponding experiments.

\begin{table}
\caption{Values of gaps $\Delta_+$ and $\Delta_-$ (in Kelvins) in two magnon branches of some relevant compounds. In the last two columns, values of $\Delta_-$ are presented which are calculated using Eq.~\eqref{d-} with constants $C$ obtained by expansion of Eq.~\eqref{an_H} to the third order in $\omega_0$ (Eq.~\eqref{c}) and from the analysis of Eq.~\eqref{an_H} without the $\omega_0$-expansion. The following parameters are used in calculations: $S=5/2$, $J_1^+=10$~K, $J_1^-=8$~K, $J_2=10$~K, and $\omega_0=1.44$~K for MnO; \cite{Pepy1974433,2pdf} $S=5/2$, $J_1^+=J_1^-=7$~K, $J_2=12$~K, and $\omega_0=0.89$~K for \ms; \cite{MnS1966} $S=5/2$, $J_1^+=J_1^-=4.9$~K, $J_2=8.1$~K, and $\omega_0=0.77$~K for \me; \cite{MnSe_Raman} $S=7/2$, $J_1^+=J_1^-=-0.384$~K, $J_2=0.626$~K, and $\omega_0=0.44$~K for EuTe; \cite{PhysRevB.81.155213} $S=7/2$, $J_1^+=J_1^-=-0.446$~K, $J_2=0.456$~K, and $\omega_0=0.53$~K for EuSe \cite{PhysRevB.81.155213}.}
\begin{tabular}{|c|c|c|c||c|c|c|}
\hline
& 
\begin{tabular}{c}
$\Delta_+$, \\
experiment \\
{}
\end{tabular}
&
\begin{tabular}{c}
$\Delta_+$, \\
theory with \\
quantum corrections \\ 
(Eqs.~\eqref{k1cor} and \eqref{d+cor})
\end{tabular} 
&
\begin{tabular}{c}
$\Delta_+$, \\
theory without \\
quantum corrections \\ 
(Eq.~\eqref{d+})
\end{tabular} 
&
\begin{tabular}{c}
$\Delta_-$, \\
experiment 
\end{tabular}
&
\begin{tabular}{c}
$\Delta_-$, \\
theory \\ 
(Eqs.~\eqref{an_H} \\
 and \eqref{d-})  
\end{tabular} 
&
\begin{tabular}{c}
$\Delta_-$, \\
theory, $\omega_0$-expansion \\ 
(Eqs.~\eqref{c} and \eqref{d-})  
\end{tabular} 
\\
\hline
MnO & 39.9 \cite{PhysRevB.13.3924,PhysRev.139.A1313,MnO_AFR} & 43.1 & 44.2 & $0.6\div1.5$ \cite{MnO_japan,PhysRev.139.A1313,PhysRevB.13.3924}
& 0.09 & 0.11 \\ 
\hline
\ms & 28.6 \cite{PhysRevB.13.3924} & 31.5 & 33.1 & 4.7 \cite{PhysRevB.13.3924} & 0.07 & 0.06 \\ 
\hline
\me & 25.9 \cite{MnSe_Raman} & 24.0 & 25.5 & $<8$ \cite{MnSe_Raman} & 0.07 & 0.05 \\ 
\hline
EuTe & 3.3 \cite{PhysRevB.1.3021,EuTe,EuTe_NMR} & 3.6 & 3.7 & $0.06\div0.17$ \cite{PhysRevB.1.3021,EuTe_NMR,demo} & 0.03 & 0.07  \\ 
\hline
EuSe & $-$ & 0.75 & 0.82 & $-$ & 0.008 & 0.015 \\ 
\hline
\end{tabular}
\end{table}

\section{Summary and conclusion}
\label{conc}

In summary, we discuss Heisenberg antiferromagnet \eqref{ham} on face-centered cubic lattice with small dipolar interaction in which the next-nearest-neighbor exchange coupling dominates over the nearest-neighbor one. 

In the limit of zero dipolar interaction, it is well known that a collinear magnetic structure is stabilized via order-by-disorder mechanism containing (111) ferromagnetic planes arranged antiferromagnetically along one of the space diagonals of the cube: quantum fluctuations stabilize the collinear arrangement of four interpenetrating AF sublattices in which the fcc lattice can be divided. Four possible (equivalent) magnetic structures of this type are described by vectors $\ko=(\pi,\pi,\pi)$, $(\pi,0,0)$, $(0,\pi,0)$, and $(0,0,\pi)$. For definiteness, we consider the magnetic structure described by vector $\ko=(\pi,\pi,\pi)$. On the mean-field level, the magnon spectrum is gapless at $\kk=\bf 0$, $(\pi,\pi,\pi)$, $(\pi,0,0)$, $(0,\pi,0)$, and $(0,0,\pi)$ (see Eqs.~\eqref{spec0} and \eqref{h2_denote_e} at $\omega_0=0$). However quantum fluctuations lead to gap \eqref{gapss} at the last three points that is related to the stabilization by quantum fluctuations of the collinear arrangement of four AF sublattices.

On the mean-field level, we find in accordance with previous results that the dipolar interaction forces spins to lie within (111) plane. We show that the dipolar interaction splits the magnon spectrum into two branches, $\epsilon_\kk^+$ and $\epsilon_\kk^-$ (see Eqs.~\eqref{specsw} and \eqref{af}). Branch $\epsilon_\kk^+$ has gap $\Delta_+$ that is related to the easy-plane anisotropy (see Eqs.~\eqref{e+}--\eqref{d+}). Another branch is gapless in the spin-wave approximation: $\epsilon_\kk^-=0$ at $\kk=\bf 0$ and $\ko$ (see Eqs.~\eqref{e-} and \eqref{Dd}). We show that the order-by-disorder mechanism leads also to the anisotropy in the easy plane \eqref{anis} which was observed experimentally in relevant compounds MnO, \ms, \me, EuTe, and EuSe. The branch $\epsilon_\kk^-$ acquires gap $\Delta_-$ \eqref{d-} in the first order in $1/S$ that is related with the in-plane anisotropy. Eqs.~\eqref{k2} and \eqref{c} give analytical expressions for the value of the in-plane anisotropy either at $\omega_0\ll|J_1^--J_1^+|$ or at $S\omega_0\ll|J_1^\pm|,J_2$ if $\omega_0\gg|J_1^--J_1^+|$. 

We compare in the Table values for gaps $\Delta_\pm$ found using expressions derived above with those obtained experimentally in MnO, \ms, \me, and EuTe. The wide scatter of values of $\Delta_-$ derived using different experimental data prevents from making definite conclusion about the origin of the in-plane anisotropy. Then, further experimental activity is needed to find more precise values of $K_2$ in these materials. Apparently, the order-by-disorder mechanism give negligible contribution to the in-plane anisotropy in compounds containing Mn whereas it can give noticeable contribution in EuTe.

We calculate also the first $1/S$ correction to the value of the easy plane anisotropy and to $\Delta_+$ that improves agreement between experimental data and the theory in all materials except for \me{ }(see Eqs.~\eqref{k1cor} and \eqref{d+cor} and the Table).

\begin{acknowledgments}

This work is supported by Russian Science Foundation (grant No.\ 14-22-00281).

\end{acknowledgments}

\bibliography{lit}

\begin{thebibliography}{40}
\expandafter\ifx\csname natexlab\endcsname\relax\def\natexlab#1{#1}\fi
\expandafter\ifx\csname bibnamefont\endcsname\relax
  \def\bibnamefont#1{#1}\fi
\expandafter\ifx\csname bibfnamefont\endcsname\relax
  \def\bibfnamefont#1{#1}\fi
\expandafter\ifx\csname citenamefont\endcsname\relax
  \def\citenamefont#1{#1}\fi
\expandafter\ifx\csname url\endcsname\relax
  \def\url#1{\texttt{#1}}\fi
\expandafter\ifx\csname urlprefix\endcsname\relax\def\urlprefix{URL }\fi
\providecommand{\bibinfo}[2]{#2}
\providecommand{\eprint}[2][]{\url{#2}}

\bibitem[{\citenamefont{Balents}(2010)}]{balents}
\bibinfo{author}{\bibfnamefont{L.}~\bibnamefont{Balents}},
  \bibinfo{journal}{Nature} \textbf{\bibinfo{volume}{464}},
  \bibinfo{pages}{199} (\bibinfo{year}{2010}).

\bibitem[{\citenamefont{Villain et~al.}(1980)\citenamefont{Villain, Bidaux,
  Carton, and Conte}}]{vill}
\bibinfo{author}{\bibfnamefont{J.}~\bibnamefont{Villain}},
  \bibinfo{author}{\bibfnamefont{R.}~\bibnamefont{Bidaux}},
  \bibinfo{author}{\bibfnamefont{J.~P.} \bibnamefont{Carton}},
  \bibnamefont{and} \bibinfo{author}{\bibfnamefont{R.}~\bibnamefont{Conte}},
  \bibinfo{journal}{J. Phys. (Paris)} \textbf{\bibinfo{volume}{41}},
  \bibinfo{pages}{1263} (\bibinfo{year}{1980}).

\bibitem[{\citenamefont{Henley}(1989)}]{henley1}
\bibinfo{author}{\bibfnamefont{C.~L.} \bibnamefont{Henley}},
  \bibinfo{journal}{Phys. Rev. Lett.} \textbf{\bibinfo{volume}{62}},
  \bibinfo{pages}{2056} (\bibinfo{year}{1989}).

\bibitem[{\citenamefont{Shender}(1982)}]{shender2}
\bibinfo{author}{\bibfnamefont{E.~F.} \bibnamefont{Shender}},
  \bibinfo{journal}{Sov. Phys. JETP} \textbf{\bibinfo{volume}{56}},
  \bibinfo{pages}{178} (\bibinfo{year}{1982}).

\bibitem[{\citenamefont{Yildirim et~al.}(1998)\citenamefont{Yildirim, Harris,
  and Shender}}]{shender}
\bibinfo{author}{\bibfnamefont{T.}~\bibnamefont{Yildirim}},
  \bibinfo{author}{\bibfnamefont{A.~B.} \bibnamefont{Harris}},
  \bibnamefont{and} \bibinfo{author}{\bibfnamefont{E.~F.}
  \bibnamefont{Shender}}, \bibinfo{journal}{Phys. Rev. B}
  \textbf{\bibinfo{volume}{58}}, \bibinfo{pages}{3144} (\bibinfo{year}{1998}).

\bibitem[{\citenamefont{Yamamoto and Nagamiya}(1972)}]{japan}
\bibinfo{author}{\bibfnamefont{Y.}~\bibnamefont{Yamamoto}} \bibnamefont{and}
  \bibinfo{author}{\bibfnamefont{T.}~\bibnamefont{Nagamiya}},
  \bibinfo{journal}{Journal of the Physical Society of Japan}
  \textbf{\bibinfo{volume}{32}}, \bibinfo{pages}{1248} (\bibinfo{year}{1972}).

\bibitem[{\citenamefont{Henley}(1987)}]{henley2}
\bibinfo{author}{\bibfnamefont{C.~L.} \bibnamefont{Henley}},
  \bibinfo{journal}{J. Appl. Phys.} \textbf{\bibinfo{volume}{61}},
  \bibinfo{pages}{3962} (\bibinfo{year}{1987}).

\bibitem[{\citenamefont{Savary et~al.}(2012)\citenamefont{Savary, Ross, Gaulin,
  Ruff, and Balents}}]{bal}
\bibinfo{author}{\bibfnamefont{L.}~\bibnamefont{Savary}},
  \bibinfo{author}{\bibfnamefont{K.~A.} \bibnamefont{Ross}},
  \bibinfo{author}{\bibfnamefont{B.~D.} \bibnamefont{Gaulin}},
  \bibinfo{author}{\bibfnamefont{J.~P.~C.} \bibnamefont{Ruff}},
  \bibnamefont{and} \bibinfo{author}{\bibfnamefont{L.}~\bibnamefont{Balents}},
  \bibinfo{journal}{Phys. Rev. Lett.} \textbf{\bibinfo{volume}{109}},
  \bibinfo{pages}{167201} (\bibinfo{year}{2012}).

\bibitem[{\citenamefont{Yosida}(1991)}]{yosida}
\bibinfo{author}{\bibfnamefont{K.}~\bibnamefont{Yosida}},
  \emph{\bibinfo{title}{Theory of Magnetism}}
  (\bibinfo{publisher}{Springer-Verlag}, \bibinfo{year}{1991}).

\bibitem[{\citenamefont{Shull et~al.}(1951)\citenamefont{Shull, Strauser, and
  Wollan}}]{neu_mno}
\bibinfo{author}{\bibfnamefont{C.~G.} \bibnamefont{Shull}},
  \bibinfo{author}{\bibfnamefont{W.~A.} \bibnamefont{Strauser}},
  \bibnamefont{and} \bibinfo{author}{\bibfnamefont{E.~O.}
  \bibnamefont{Wollan}}, \bibinfo{journal}{Phys. Rev.}
  \textbf{\bibinfo{volume}{83}}, \bibinfo{pages}{333} (\bibinfo{year}{1951}).

\bibitem[{\citenamefont{Goodwin et~al.}(2007)\citenamefont{Goodwin, Dove,
  Tucker, and Keen}}]{good}
\bibinfo{author}{\bibfnamefont{A.~L.} \bibnamefont{Goodwin}},
  \bibinfo{author}{\bibfnamefont{M.~T.} \bibnamefont{Dove}},
  \bibinfo{author}{\bibfnamefont{M.~G.} \bibnamefont{Tucker}},
  \bibnamefont{and} \bibinfo{author}{\bibfnamefont{D.~A.} \bibnamefont{Keen}},
  \bibinfo{journal}{Phys. Rev. B} \textbf{\bibinfo{volume}{75}},
  \bibinfo{pages}{075423} (\bibinfo{year}{2007}).

\bibitem[{\citenamefont{Corliss et~al.}(1956)\citenamefont{Corliss, Elliott,
  and Hastings}}]{neu_mns}
\bibinfo{author}{\bibfnamefont{L.}~\bibnamefont{Corliss}},
  \bibinfo{author}{\bibfnamefont{N.}~\bibnamefont{Elliott}}, \bibnamefont{and}
  \bibinfo{author}{\bibfnamefont{J.}~\bibnamefont{Hastings}},
  \bibinfo{journal}{Phys. Rev.} \textbf{\bibinfo{volume}{104}},
  \bibinfo{pages}{924} (\bibinfo{year}{1956}).

\bibitem[{\citenamefont{Will et~al.}(1963)\citenamefont{Will, Pickart, Alperin,
  and Nathans}}]{eute_neu}
\bibinfo{author}{\bibfnamefont{G.}~\bibnamefont{Will}},
  \bibinfo{author}{\bibfnamefont{S.~J.} \bibnamefont{Pickart}},
  \bibinfo{author}{\bibfnamefont{H.~A.} \bibnamefont{Alperin}},
  \bibnamefont{and} \bibinfo{author}{\bibfnamefont{R.}~\bibnamefont{Nathans}},
  \bibinfo{journal}{J. Phys. Chem. Solids} \textbf{\bibinfo{volume}{24}},
  \bibinfo{pages}{1679} (\bibinfo{year}{1963}).

\bibitem[{\citenamefont{{Griessen} et~al.}(1971)\citenamefont{{Griessen},
  {Landolt}, and {Ott}}}]{Griessen19712219}
\bibinfo{author}{\bibfnamefont{R.}~\bibnamefont{{Griessen}}},
  \bibinfo{author}{\bibfnamefont{M.}~\bibnamefont{{Landolt}}},
  \bibnamefont{and} \bibinfo{author}{\bibfnamefont{H.~R.} \bibnamefont{{Ott}}},
  \bibinfo{journal}{Solid State Communications} \textbf{\bibinfo{volume}{9}},
  \bibinfo{pages}{2219} (\bibinfo{year}{1971}).

\bibitem[{\citenamefont{S\"ollinger et~al.}(2010)\citenamefont{S\"ollinger,
  Heiss, Lechner, Rumpf, Granitzer, Krenn, and
  Springholz}}]{PhysRevB.81.155213}
\bibinfo{author}{\bibfnamefont{W.}~\bibnamefont{S\"ollinger}},
  \bibinfo{author}{\bibfnamefont{W.}~\bibnamefont{Heiss}},
  \bibinfo{author}{\bibfnamefont{R.~T.} \bibnamefont{Lechner}},
  \bibinfo{author}{\bibfnamefont{K.}~\bibnamefont{Rumpf}},
  \bibinfo{author}{\bibfnamefont{P.}~\bibnamefont{Granitzer}},
  \bibinfo{author}{\bibfnamefont{H.}~\bibnamefont{Krenn}}, \bibnamefont{and}
  \bibinfo{author}{\bibfnamefont{G.}~\bibnamefont{Springholz}},
  \bibinfo{journal}{Phys. Rev. B} \textbf{\bibinfo{volume}{81}},
  \bibinfo{pages}{155213} (\bibinfo{year}{2010}).

\bibitem[{\citenamefont{Cohen and Keffer}(1955)}]{cohen}
\bibinfo{author}{\bibfnamefont{M.~H.} \bibnamefont{Cohen}} \bibnamefont{and}
  \bibinfo{author}{\bibfnamefont{F.}~\bibnamefont{Keffer}},
  \bibinfo{journal}{Phys. Rev.} \textbf{\bibinfo{volume}{99}},
  \bibinfo{pages}{1128} (\bibinfo{year}{1955}).

\bibitem[{\citenamefont{Kaplan}(1954)}]{kaplan}
\bibinfo{author}{\bibfnamefont{J.~I.} \bibnamefont{Kaplan}},
  \bibinfo{journal}{The Journal of Chemical Physics}
  \textbf{\bibinfo{volume}{22}}, \bibinfo{pages}{1709} (\bibinfo{year}{1954}).

\bibitem[{\citenamefont{Keffer and O'Sullivan}(1957)}]{1957}
\bibinfo{author}{\bibfnamefont{F.}~\bibnamefont{Keffer}} \bibnamefont{and}
  \bibinfo{author}{\bibfnamefont{W.}~\bibnamefont{O'Sullivan}},
  \bibinfo{journal}{Phys. Rev.} \textbf{\bibinfo{volume}{108}},
  \bibinfo{pages}{637} (\bibinfo{year}{1957}).

\bibitem[{\citenamefont{Rodbell et~al.}(1963)\citenamefont{Rodbell, Jacobs,
  Owen, and Harris}}]{bic1}
\bibinfo{author}{\bibfnamefont{D.~S.} \bibnamefont{Rodbell}},
  \bibinfo{author}{\bibfnamefont{I.~S.} \bibnamefont{Jacobs}},
  \bibinfo{author}{\bibfnamefont{J.}~\bibnamefont{Owen}}, \bibnamefont{and}
  \bibinfo{author}{\bibfnamefont{E.~A.} \bibnamefont{Harris}},
  \bibinfo{journal}{Phys. Rev. Lett.} \textbf{\bibinfo{volume}{11}},
  \bibinfo{pages}{10} (\bibinfo{year}{1963}).

\bibitem[{\citenamefont{D\'{\i}az et~al.}(2008)\citenamefont{D\'{\i}az,
  Granado, Abramof, Rappl, Chitta, and Henriques}}]{bic2}
\bibinfo{author}{\bibfnamefont{B.}~\bibnamefont{D\'{\i}az}},
  \bibinfo{author}{\bibfnamefont{E.}~\bibnamefont{Granado}},
  \bibinfo{author}{\bibfnamefont{E.}~\bibnamefont{Abramof}},
  \bibinfo{author}{\bibfnamefont{P.~H.~O.} \bibnamefont{Rappl}},
  \bibinfo{author}{\bibfnamefont{V.~A.} \bibnamefont{Chitta}},
  \bibnamefont{and} \bibinfo{author}{\bibfnamefont{A.~B.}
  \bibnamefont{Henriques}}, \bibinfo{journal}{Phys. Rev. B}
  \textbf{\bibinfo{volume}{78}}, \bibinfo{pages}{134423}
  (\bibinfo{year}{2008}).

\bibitem[{\citenamefont{Harris and Owen}(1963)}]{bic3}
\bibinfo{author}{\bibfnamefont{E.~A.} \bibnamefont{Harris}} \bibnamefont{and}
  \bibinfo{author}{\bibfnamefont{J.}~\bibnamefont{Owen}},
  \bibinfo{journal}{Phys. Rev. Lett.} \textbf{\bibinfo{volume}{11}},
  \bibinfo{pages}{9} (\bibinfo{year}{1963}).

\bibitem[{\citenamefont{Pepy}(1974)}]{Pepy1974433}
\bibinfo{author}{\bibfnamefont{G.}~\bibnamefont{Pepy}},
  \bibinfo{journal}{Journal of Physics and Chemistry of Solids}
  \textbf{\bibinfo{volume}{35}}, \bibinfo{pages}{433} (\bibinfo{year}{1974}).

\bibitem[{\citenamefont{Morosin}(1970)}]{PhysRevB.1.236}
\bibinfo{author}{\bibfnamefont{B.}~\bibnamefont{Morosin}},
  \bibinfo{journal}{Phys. Rev. B} \textbf{\bibinfo{volume}{1}},
  \bibinfo{pages}{236} (\bibinfo{year}{1970}).

\bibitem[{\citenamefont{Zinn}(1976)}]{Zinn197623}
\bibinfo{author}{\bibfnamefont{W.}~\bibnamefont{Zinn}},
  \bibinfo{journal}{Journal of Magnetism and Magnetic Materials}
  \textbf{\bibinfo{volume}{3}}, \bibinfo{pages}{23 } (\bibinfo{year}{1976}).

\bibitem[{\citenamefont{Batalov and Syromyatnikov}(2015)}]{our}
\bibinfo{author}{\bibfnamefont{L.~A.} \bibnamefont{Batalov}} \bibnamefont{and}
  \bibinfo{author}{\bibfnamefont{A.~V.} \bibnamefont{Syromyatnikov}},
  \bibinfo{journal}{Phys. Rev. B} \textbf{\bibinfo{volume}{91}},
  \bibinfo{pages}{224432} (\bibinfo{year}{2015}).

\bibitem[{\citenamefont{Syromyatnikov}(2006)}]{syromyat3d1}
\bibinfo{author}{\bibfnamefont{A.~V.} \bibnamefont{Syromyatnikov}},
  \bibinfo{journal}{Phys. Rev. B} \textbf{\bibinfo{volume}{74}},
  \bibinfo{pages}{014435} (\bibinfo{year}{2006}).

\bibitem[{\citenamefont{Syromyatnikov}(2008)}]{syromyat2d}
\bibinfo{author}{\bibfnamefont{A.~V.} \bibnamefont{Syromyatnikov}},
  \bibinfo{journal}{Phys. Rev. B} \textbf{\bibinfo{volume}{77}},
  \bibinfo{pages}{144433} (\bibinfo{year}{2008}).

\bibitem[{\citenamefont{Syromyatnikov}(2010)}]{syromyat3d}
\bibinfo{author}{\bibfnamefont{A.~V.} \bibnamefont{Syromyatnikov}},
  \bibinfo{journal}{Phys. Rev. B} \textbf{\bibinfo{volume}{82}},
  \bibinfo{pages}{024432} (\bibinfo{year}{2010}).

\bibitem[{\citenamefont{Hihara and Kawakami}(1988)}]{EuTe_NMR}
\bibinfo{author}{\bibfnamefont{T.}~\bibnamefont{Hihara}} \bibnamefont{and}
  \bibinfo{author}{\bibfnamefont{M.}~\bibnamefont{Kawakami}},
  \bibinfo{journal}{Journal of the Physical Society of Japan}
  \textbf{\bibinfo{volume}{57}}, \bibinfo{pages}{1094} (\bibinfo{year}{1988}).

\bibitem[{\citenamefont{Battles and Everett}(1970)}]{PhysRevB.1.3021}
\bibinfo{author}{\bibfnamefont{J.~W.} \bibnamefont{Battles}} \bibnamefont{and}
  \bibinfo{author}{\bibfnamefont{G.~E.} \bibnamefont{Everett}},
  \bibinfo{journal}{Phys. Rev. B} \textbf{\bibinfo{volume}{1}},
  \bibinfo{pages}{3021} (\bibinfo{year}{1970}).

\bibitem[{\citenamefont{Oliveira et~al.}(1972)\citenamefont{Oliveira, Foner,
  Shapira, and Reed}}]{PhysRevB.5.2634}
\bibinfo{author}{\bibfnamefont{N.~F.} \bibnamefont{Oliveira}},
  \bibinfo{author}{\bibfnamefont{S.}~\bibnamefont{Foner}},
  \bibinfo{author}{\bibfnamefont{Y.}~\bibnamefont{Shapira}}, \bibnamefont{and}
  \bibinfo{author}{\bibfnamefont{T.~B.} \bibnamefont{Reed}},
  \bibinfo{journal}{Phys. Rev. B} \textbf{\bibinfo{volume}{5}},
  \bibinfo{pages}{2634} (\bibinfo{year}{1972}).

\bibitem[{\citenamefont{{Demokritov} et~al.}(1986)\citenamefont{{Demokritov},
  {Kre{\v i}nes}, and {Kudinov}}}]{demo}
\bibinfo{author}{\bibfnamefont{S.~O.} \bibnamefont{{Demokritov}}},
  \bibinfo{author}{\bibfnamefont{N.~M.} \bibnamefont{{Kre{\v i}nes}}},
  \bibnamefont{and} \bibinfo{author}{\bibfnamefont{V.~I.}
  \bibnamefont{{Kudinov}}}, \bibinfo{journal}{JETP Lett.}
  \textbf{\bibinfo{volume}{43}}, \bibinfo{pages}{403} (\bibinfo{year}{1986}).

\bibitem[{\citenamefont{Kohgi et~al.}(1972)\citenamefont{Kohgi, Ishikawa, and
  Endoh}}]{2pdf}
\bibinfo{author}{\bibfnamefont{M.}~\bibnamefont{Kohgi}},
  \bibinfo{author}{\bibfnamefont{Y.}~\bibnamefont{Ishikawa}}, \bibnamefont{and}
  \bibinfo{author}{\bibfnamefont{Y.}~\bibnamefont{Endoh}},
  \bibinfo{journal}{Solid State Communications} \textbf{\bibinfo{volume}{11}},
  \bibinfo{pages}{391 } (\bibinfo{year}{1972}).

\bibitem[{\citenamefont{Lines and Jones}(1966)}]{MnS1966}
\bibinfo{author}{\bibfnamefont{M.~E.} \bibnamefont{Lines}} \bibnamefont{and}
  \bibinfo{author}{\bibfnamefont{E.~D.} \bibnamefont{Jones}},
  \bibinfo{journal}{Phys. Rev.} \textbf{\bibinfo{volume}{141}},
  \bibinfo{pages}{525} (\bibinfo{year}{1966}).

\bibitem[{\citenamefont{Milutinovi\ifmmode~\acute{c}\else \'{c}\fi{}
  et~al.}(2002)\citenamefont{Milutinovi\ifmmode~\acute{c}\else \'{c}\fi{},
  Tomi\ifmmode~\acute{c}\else \'{c}\fi{}, Devi\ifmmode~\acute{c}\else
  \'{c}\fi{}, Milutinovi\ifmmode~\acute{c}\else \'{c}\fi{}, and
  Popovi\ifmmode~\acute{c}\else \'{c}\fi{}}}]{MnSe_Raman}
\bibinfo{author}{\bibfnamefont{A.}~\bibnamefont{Milutinovi\ifmmode~\acute{c}\else
  \'{c}\fi{}}},
  \bibinfo{author}{\bibfnamefont{N.}~\bibnamefont{Tomi\ifmmode~\acute{c}\else
  \'{c}\fi{}}},
  \bibinfo{author}{\bibfnamefont{S.}~\bibnamefont{Devi\ifmmode~\acute{c}\else
  \'{c}\fi{}}},
  \bibinfo{author}{\bibfnamefont{P.}~\bibnamefont{Milutinovi\ifmmode~\acute{c}\else
  \'{c}\fi{}}}, \bibnamefont{and} \bibinfo{author}{\bibfnamefont{Z.~V.}
  \bibnamefont{Popovi\ifmmode~\acute{c}\else \'{c}\fi{}}},
  \bibinfo{journal}{Phys. Rev. B} \textbf{\bibinfo{volume}{66}},
  \bibinfo{pages}{012302} (\bibinfo{year}{2002}).

\bibitem[{\citenamefont{Chou and Fan}(1976)}]{PhysRevB.13.3924}
\bibinfo{author}{\bibfnamefont{H.-h.} \bibnamefont{Chou}} \bibnamefont{and}
  \bibinfo{author}{\bibfnamefont{H.~Y.} \bibnamefont{Fan}},
  \bibinfo{journal}{Phys. Rev. B} \textbf{\bibinfo{volume}{13}},
  \bibinfo{pages}{3924} (\bibinfo{year}{1976}).

\bibitem[{\citenamefont{Lines and Jones}(1965)}]{PhysRev.139.A1313}
\bibinfo{author}{\bibfnamefont{M.~E.} \bibnamefont{Lines}} \bibnamefont{and}
  \bibinfo{author}{\bibfnamefont{E.~D.} \bibnamefont{Jones}},
  \bibinfo{journal}{Phys. Rev.} \textbf{\bibinfo{volume}{139}},
  \bibinfo{pages}{A1313} (\bibinfo{year}{1965}).

\bibitem[{\citenamefont{Hughes}(1971)}]{MnO_AFR}
\bibinfo{author}{\bibfnamefont{A.~E.} \bibnamefont{Hughes}},
  \bibinfo{journal}{Phys. Rev. B} \textbf{\bibinfo{volume}{3}},
  \bibinfo{pages}{877} (\bibinfo{year}{1971}).

\bibitem[{\citenamefont{Uchida et~al.}(1960)\citenamefont{Uchida, Kondoh,
  Nakazumi, and Nagamiya}}]{MnO_japan}
\bibinfo{author}{\bibfnamefont{E.}~\bibnamefont{Uchida}},
  \bibinfo{author}{\bibfnamefont{H.}~\bibnamefont{Kondoh}},
  \bibinfo{author}{\bibfnamefont{Y.}~\bibnamefont{Nakazumi}}, \bibnamefont{and}
  \bibinfo{author}{\bibfnamefont{T.}~\bibnamefont{Nagamiya}},
  \bibinfo{journal}{Journal of the Physical Society of Japan}
  \textbf{\bibinfo{volume}{15}}, \bibinfo{pages}{466} (\bibinfo{year}{1960}).

\bibitem[{\citenamefont{Masset and Callaway}(1970)}]{EuTe}
\bibinfo{author}{\bibfnamefont{F.}~\bibnamefont{Masset}} \bibnamefont{and}
  \bibinfo{author}{\bibfnamefont{J.}~\bibnamefont{Callaway}},
  \bibinfo{journal}{Phys. Rev. B} \textbf{\bibinfo{volume}{2}},
  \bibinfo{pages}{3657} (\bibinfo{year}{1970}).

\end{thebibliography}

\end{document}